\documentclass[pra]{revtex4}
\usepackage{graphicx}

\begin{document}

\title{A Realistic Radiative Fermion Mass Hierarchy in Non-supersymmetric $SO(10)$ }

\author{{\bf S.M. Barr and Almas Khan} \\
Bartol Research Institute \\ University of Delaware \\
Newark, Delaware 19716}

\date{\today}

\begin{abstract}
A non-supersymmetric grand unified theory can exhibit a ``radiative
fermion mass hierarchy", in which the heavier quarks and leptons
get mass at tree level and the lighter ones get mass from loop diagrams.
Recently the first predictive model of this type was proposed. Here it
is analyzed numerically and it is shown to give an excellent fit to
the quark and lepton masses and mixings, including the CP phase violating
phase $\delta_{CKM}$. A relation between the neutrino angle
$\theta_{13}$ and the atmospheric neutrino angle is obtained.
\end{abstract}

\maketitle

\section{Introduction}

The masses of the known quarks and leptons exhibit a large
hierarchy.  This has suggested to many theorists \cite{rad}
that the light fermion mass hierarchy could be ``radiative", i.e.
that the lightest fermions get mass from loop diagrams, while the
heaviest get mass at tree level. In the early 1980's several papers
showed that such an idea can be implemented naturally in the context
of non-supersymmetric grand unified theories (GUTs) \cite{radgut}.
In models of this type, the radiative masses come from loop diagrams
containing virtual GUT-scale particles. That is why such
models must be non-supersymmetric:
otherwise, the loops would be suppressed by
$O(M_{SUSY}^2/M_{GUT}^2)$ due to the non-renormalization theorems of
supersymmetry.

In a recent paper \cite{so10rad} a very simple non-supersymmetric
$SO(10)$ model with a radiative hierarchy was proposed. One thing
that allows this model to be so simple is precisely that its
hierarchy is radiative. The point is that terms have to
exist in the lagrangian corresponding to the larger elements of the quark
and lepton mass matrices, but not to the smallest elements, since they
arise automatically from loops. The simplification can be seen by
comparing the model of \cite{so10rad} to the supersymmetric
$SO(10)$ model on which
it was based, which had a non-radiative hierarchy \cite{bb2002}.
That earlier model had a somewhat larger particle content and more
Yukawa terms.

Models with radiative hierarchies are also simpler in another way:
in them it is not necessary to introduce {\it ad hoc} very small
dimensionless parameters to account for the fermion mass
hierarchies, since they are automatically accounted for by the loop
factors $1/16 \pi^2$. Despite radiative hierarchy models being able to have a
simpler structure, one might think they would be less predictive,
since loop diagrams tend to depend on many parameters. However, the
model proposed in \cite{so10rad} shows that this need not be the
case. In that paper it was shown that the model gives a
qualitatively realistic pattern of quark and lepton masses and
mixings with only 9 parameters.

While the non-supersymmetric model proposed in \cite{so10rad} is
economical and qualitatively realistic, the analysis in that paper
was not sufficient to establish that it is realistic quantitatively.
In particular, several issues were not addressed. First, it was not
specified what the sequence and scales of breaking were of $SO(10)$
down to the Standard Model group $G_{SM}$ (which must, of course be
consistent with proton decay bounds and unification of gauge
couplings). Unless that is done, the renormalization-group running
of the quark and lepton masses needed for a global fit of parameters
cannot be performed. Second, the forms of the mass matrices given
\cite{so10rad} were derived under the assumption that certain
$SU(5)$-breaking effects could be ignored. However, as will be seen,
this assumption is not necessarily consistent with the pattern of
$SO(10)$ breaking that needs to be assumed in order to satisfy the
constraints of gauge coupling unification and proton decay.
$SU(5)$-breaking effects will turn out to modify significantly the
forms of the quark and lepton mass matrices given in \cite{so10rad}.
Third, the $m_b/m_{\tau}$ ratio is problematic in the version of the
model discussed in \cite{so10rad}. In that model, because certain
$SU(5)$-breaking effects were treated as negligible, the classic
prediction $m_b^0 \cong m_{\tau}^0$ was obtained. (A superscript `0'
indicates throughout this paper quantities evaluated at the GUT scale.)
This is well-known to give a fairly good fit in supersymmetric
models for certain values of $\tan \beta$ \cite{btaususy}; but in
non-supersymmetric models it results in a prediction of
$m_b/m_{\tau}$ at low energies that is typically too large by at
least $30\%$ \cite{btaunonsusy}. Fourth, the ratio $m_s/m_b$ is
predicted in the version of the model given in to have the
Georgi-Jarlskog value $\frac{1}{3} m_{\mu}/m_{\tau}$ at the GUT
scale \cite{gj}. However, lattice calculations \cite{lattice} have
suggested that $m_s$ is significantly smaller than previous
estimates of it, and the best fit value is now somewhat smaller than
the Georgi-Jarlskog prediction.

In this paper, we address all these issues. The paper is organized
as follows. In section 2, the $SO(10)$ model of \cite{so10rad} is
reviewed, and it is explained how both the tree-level and radiative
contributions to the mass matrices arise, and why the resulting
forms give a good qualitative description of the pattern of quark
and lepton masses and mixings. In section 3, a breaking of $SO(10)$
down to the Standard Model consistent with gauge coupling
unification and proton decay bounds is specified. In section 4, the
effect of this pattern of symmetry breaking on the quark and lepton
mass matrices is discussed and it is shown that forms somewhat
different from those given in \cite{so10rad} result. In section 5,
the results of a global numerical fit to the quark and lepton masses
and mixings is given. An excellent fit is found to the quark and lepton
masses and mixings, including the CP phase $\delta_{CKM}$. A relation
between the neutrino angle $\theta_{13}$ and
the atmospheric neutrino angle is obtained.

\section{The model}

The model proposed in \cite{so10rad}, whose predictions we analyze
in detail in this paper, is a non-supersymmetric with unified group $SO(10)$.
In it the tree-level mass matrices of quarks and charged leptons are
generated by only
three effective Yukawa operators

\begin{equation}
\begin{array}{l}
O_1 = {\bf 16}_3 {\bf 16}_3 {\bf 10}_H  \\
O_2 = {\bf 16}_2 {\bf 16}_3 {\bf 10}_H {\bf 45}_H/M_{GUT} \\
O_3 = (c_i {\bf 16}_i {\bf 16}_{iH}) ({\bf 16}_3 {\bf 16}'_H)/M_{GUT} \;\;\;, i=1,2 \\
\end{array}
\end{equation}

\noindent
In $O_3$, the factors in parentheses are contracted into ${\bf 10}$'s of $SO(10)$.
The loop-level elements in the mass matrices arise very simply from the tree-level
elements, as will be seen later. The three operators given in Eq. (1) do the following
things: $O_1$ gives the 33 elements of the mass matrices, i.e. the masses of the third
family. $O_2$ and $O_3$ generate the masses of the second family and its mixing with
the third family (i.e. $V_{cb}$ and $\theta_{atm}$), and also $\theta_{sol}$. The
masses of the first family and its mixings come from loops.

The operators in Eq. (1) come from integrating out some ``extra" vectorlike fermion
multiplets, consisting of an $SO(10)$ vector and a spinor-antispinor pair.
Thus the complete fermionic content of the model comprises
the following (left-handed) multiplets:
${\bf 16}_{i = 1,2,3} + \;\;
({\bf 16} + \overline{{\bf 16}} + {\bf 10})$. The Dirac mass
matrices of the up-type quarks, down-type quarks, charged leptons, and neutrinos
(denoted by
$M_U$, $M_D$, $M_L$, and $M_N$, respectively) arise from the following set of
Yukawa terms in the lagrangian:

\begin{equation}
\begin{array}{ccl}
{\cal L}_{Yuk} & = & M_{16} ( \overline{{\bf 16}} \; {\bf 16}) +
M_{10} ( {\bf 10} \; {\bf 10}).
\\ & & \\
& + & a( \overline{{\bf 16}} \; {\bf 16}_3 ) {\bf 45}_H + \sum_{i=1,2} c_i (
{\bf 10} \;
{\bf 16}_i) {\bf 16}_{iH} \\ & & \\
& + & h_{33} ( {\bf 16}_3 {\bf 16}_3 ) {\bf 10}_H + h_2 ( {\bf 16} \; {\bf 16}_2)
{\bf 10}_H + h_3 ( {\bf 10} \; {\bf 16}_3)
{\bf 16}'_H \\ & & \\
& + & h ( {\bf 16} \; {\bf 16} ) {\bf 10}'_H. \end{array}
\end{equation}

\noindent It is shown in \cite{so10rad} that this form of the Yukawa
interactions is the most general allowed by a certain simple $U(1)$ flavor
symmetry, which will be denoted by $U(1)_F$. The terms on the first
line of Eq. (2) are the $O(M_{GUT})$ masses of the extra fermion
multiplets; the terms on the second line contribute $O(M_{GUT})$
masses that mix those extra fermions with the three chiral families
${\bf 16}_i$; the terms on the third line generate the weak-scale
$SU(2)_L \times U(1)_Y$-breaking masses; and the last term is needed
to give radiative masses to the first family. Higgs multiplets are
denoted by the subscript $H$. The Higgs fields ${\bf 16}_{iH}$
obtain vacuum expectation values (VEV) in the ${\bf 1} ({\bf 16})$
direction. (The expression ${\bf p}({\bf q})$ stands for a ${\bf p}$
multiplet of $SU(5)$ contained in a ${\bf q}$ multiplet of
$SO(10)$.) The adjoint Higgs field ${\bf 45}_H$ is assumed to obtain
a VEV that is proportional to the $SO(10)$ generator $B-L$ (i.e.
baryon number minus lepton number).

The electroweak gauge symmetry $SU(2)_L \otimes U(1)_Y$ is
spontaneously broken by the Higgs multiplets denoted ${\bf 10}_H$,
${\bf 10}'_H$, and ${\bf 16}'_H$ in Eq. (1), and, more specifically,
by the neutral components of the $Y/2= -1/2$ doublets contained in
$\overline{{\bf 5}}({\bf 10}_H)$, $\overline{{\bf 5}}({\bf 10}'_H)$,
and $\overline{{\bf 5}}({\bf 16}'_H)$, and the neutral components of
the $Y/2 = +1/2$ doublets contained in ${\bf 5}({\bf 10}_H)$ and
${\bf 5}({\bf 10}'_H)$. Of course, in the low-energy effective
theory, which is just the Standard Model, there is only one Higgs
doublet, which is some linear combination of these doublets (and
their hermitian conjugates).

According to \cite{so10rad}, the mass matrices that result from the terms in Eq. (2) have the form

\begin{equation}
\begin{array}{ll}
M_U = \left( \begin{array}{ccc} 0 & 0 & 0 \\
0 & 0 & \frac{\epsilon}{3} \\ 0 & - \frac{\epsilon}{3} & 1
\end{array} \right) \; m_U, \;\;\; & M_D = \left( \begin{array}{ccc}
0 & 0 & \delta_{g1} \\ 0 & \delta_H & \frac{\epsilon}{3} + \delta_{g2} \\
C_1 & C_2 - \frac{\epsilon}{3} & 1 \end{array} \right) \; m_D, \\ & \\
M_N = \left( \begin{array}{ccc} 0 & 0 & 0 \\
0 & 0 & - \epsilon \\ 0 & \epsilon & 1 \end{array}
\right) \; m_U, \;\;\; & M_L = \left( \begin{array}{ccc}
0 & 0 & C_1 \\ 0 & \delta_H & C_2 - \epsilon \\
\delta_{g1} & \epsilon + \delta_{g2} & 1 \end{array} \right) \; m_D,
\end{array}
\end{equation}

\noindent
where $m_U \equiv h_{33} \langle {\bf 5}({\bf 10}_H) \rangle$ and
$m_D \equiv h_{33} \langle \overline{{\bf 5}}({\bf 10}_H) \rangle$.
(It will be seen in section 4 that GUT-symmetry-breaking effects
modify these forms somewhat.)
The convention here is that the mass matrices are multiplied from the left by
the left-handed fermions and from the right by the right-handed fermions.

The 33 elements of the mass matrices in Eq. (3) come simply from
the term $h_{33} ( {\bf 16}_3 \; {\bf 16}_3 ) {\bf 10}_H$, as is
usually the case in $SO(10)$ models \cite{33so10}. (This is just the operator
$O_1$ in Eq. (1).)

The contributions to the 23 and 32 elements denoted by $\epsilon$
come from integrating out the family-antifamily pair $\overline{{\bf
16}} + {\bf 16}$. The antifamily $\overline{{\bf 16}}$ appears in
two mass terms from Eq. (2), which can be combined as follows:
$\overline{{\bf 16}} (M_{16} {\bf 16} + a \langle {\bf 45}_H \rangle
{\bf 16}_3)$. These terms have the effect of mixing the ${\bf 16}$
with the ${\bf 16}_3$. One linear combination of ${\bf 16}$ and ${\bf 16}_3$
obtains an $O(M_{GUT})$ mass,
while the orthogonal combination (denoted by the index $3'$) remains light. (From now
on, primed indices will be used to denote the light families that
remain after the superheavy fermions have been integrated out.)
Thus, the ${\bf 16}$ with no index has some of the third light
family mixed in with it; and the amount of this mixing is
proportional to the VEV $\langle {\bf 45}_H \rangle$. As a result,
the term $h_2 ( {\bf 16} \; {\bf 16}_2) {\bf 10}_H$ from Eq. (2)
leads to an effective operator of the form $({\bf 16}_{3'} {\bf
16}_{2'}) {\bf 10}_H {\bf 45}_H/M_{GUT}$, which is just
the operator $O_2$ of Eq. (1), which in turn produces the
contributions denoted in Eq. (3) by $\epsilon$. Since $\langle {\bf
45}_H \rangle \propto B-L$, the $\epsilon$ contributions are $1/3$
times as large for the quarks as for the leptons.

The elements denoted by $C_1$ and $C_2$ arise in a similar fashion
by integrating out the $SO(10)$-vector multiplet of quarks and
leptons, ${\bf 10}$. This multiplet contains a $\overline{{\bf 5}} +
{\bf 5}$ of $SU(5)$. The ${\bf 5}({\bf 10})$ appears in several mass
terms from Eq. (2), which can be combined as ${\bf 5}({\bf 10}) [
M_{10} \overline{{\bf 5}} ({\bf 10}) + \sum_{i=1,2} c_i \langle {\bf
1}({\bf 16}_{iH}) \rangle \overline{{\bf 5}} ({\bf 16}_i) ]$. These
terms have the effect of mixing the $\overline{{\bf 5}}({\bf 10})$
with the $\overline{{\bf 5}} ({\bf 16}_1)$ and $\overline{{\bf
5}}({\bf 16}_2)$. One linear combination of these $\overline{{\bf
5}}$'s obtains an $O(M_{GUT})$ mass, while the two orthogonal linear
combinations are in the light families and denoted $\overline{{\bf
5}}_{1'}$ and $\overline{{\bf 5}}_{2'}$. Consequently, the
$\overline{{\bf 5}}({\bf 10})$ has mixed in with it some of
$\overline{{\bf 5}}_{1'}$ and $\overline{{\bf 5}}_{2'}$. That means
that the term $h_3 ({\bf 10} \; {\bf 16}_3 ) {\bf 16}'_H$ in Eq. (2)
leads to effective mass terms of the form $(C_1 \overline{{\bf
5}}_{1'} + C_2 \overline{{\bf 5}}_{2'}) {\bf 10}_{3'} m_D$. This is just
the operator $O_3$ of Eq. (1) and
gives the terms denoted by the $C_i$ in Eq. (3). These contributions
appear only in $M_L$ and $M_D$, because $\overline{{\bf 5}}$'s of
$SU(5)$ contain only charged leptons and down-type quarks. In both
\cite{bb2002} and \cite{so10rad} the $M_{10}$ was assumed to be an
explicit (and therefore $SU(5)$-invariant) mass, and therefore the
same $C_i$ appear in both $M_L$ and $M_D$.

At this point it should be noted that the expressions for the quark
and lepton mass matrices given in Eq. (3) are approximate. The exact
expressions involve factors, such as $1/\sqrt{1 + (a \langle {\bf
45}_H \rangle/M_{16})^2}$ and $1/\sqrt{1 + (\sum_i c_i \langle {\bf
16}_{iH} \rangle/M_{10})^2}$, which are essentially just the cosines
of angles describing the mixing between the extra fermions ${\bf 16}
+ \overline{{\bf 16}} + {\bf 10}$ and the three chiral families
${\bf 16}_i$. If these mixing angles are small, their cosines are
very close to one, and the mass matrices become insensitive to their
values. This is an assumption that we make here (as in
\cite{so10rad}), as it reduces the number of parameters. However,
there is no {\it a priori} reason to assume that these angles are
extremely small. (Indeed, if they vanished, so would $\epsilon$.) If
one of these angles were of order $0.25$ radians, say, it would give $3\%$
corrections to some of the elements of the mass matrices.

The elements denoted by $\delta_{gi}$ and $\delta_H$ in Eq. (3) are
necessary to make the mass matrices $M_L$ and $M_D$ be of rank 3
rather than rank 2, and so generate masses and mixings for the first
family. As will be seen, in order to fit the first family masses and
mixings these $\delta$ are must be of order $10^{-2}$, whereas the
other parameters appearing inside in the mass matrices in Eq. (3)
turn out to be of order 1 (or, in the case of $\epsilon$, about
$0.19$). In \cite{bb2002}, additional vectorlike quark and lepton
fields besides those in Eq. (2) had to be introduced in order to generate
these small $\delta$'s. In \cite{so10rad}, however, it was noted that that the
terms in Eq. (2) are enough to generate the $\delta$ terms
automatically by one-loop diagrams and also to explain why they are
of order $10^{-2}$.

The $\delta_{gi}$ terms are given by the one-gauge-boson-loop
diagram shown in Fig. 1(a). The gauge boson in this diagram is in a
${\bf 10}$ of $SU(5)$ (of course, it is in the adjoint ${\bf 45}$ of
$SO(10)$), so that it turns ${\bf 10}$'s of $SU(5)$ into
$\overline{{\bf 5}}$'s and {\it vice versa}. That means that the
small $\delta_{gi}$ elements that couple ${\bf 10}_i$ to
$\overline{{\bf 5}}_3$ (namely $(M_D)_{i3}$ and $(M_L)_{3i}$) come
from the large $C_i$ elements that couple $\overline{{\bf 5}}_i$
to ${\bf 10}_3$
(namely $(M_L)_{i3}$ and $(M_D)_{3i}$) So
$\delta_{gi} \propto C_i$. These diagrams were evaluated in
\cite{so10rad} neglecting certain $SU(5)$-breaking effects, giving
the result that the same $\delta_{gi}$ appear in $M_L$ and $M_D$, as
given in Eq. (3).

\begin{center}
\begin{picture}(360,140)
\thicklines \put(60,60){\vector(1,0){30}}
\put(90,60){\line(1,0){30}} \put(120,60){\vector(1,0){30}}
\put(150,60){\line(1,0){60}} \put(240,60){\vector(-1,0){30}}
\put(240,60){\line(1,0){30}} \put(300,60){\vector(-1,0){30}}
\put(180,60){\oval(120,80)[t]} \put(180,60){\circle*{20}}
\put(210,100){\vector(-1,0){30}} \put(70,45){${\bf 10}({\bf 16}_i)$}
\put(130,45){$\overline{{\bf 5}}({\bf 16}_i)$} \put(195,45){${\bf
10}({\bf 16}_3)$} \put(250,45){$\overline{{\bf 5}}({\bf 16}_3)$}
\put(160,110){${\bf 10}({\bf 45}_g)$} \put(175,40){$C_i$}
\put(160,20){{\bf (a)}}
\end{picture}
\end{center}

\begin{center}
\begin{picture}(360,140)
\thicklines \put(30,70){\vector(1,0){25}}
\put(55,70){\line(1,0){25}} \put(80,70){\vector(1,0){25}}
\put(105,70){\line(1,0){50}} \put(180,70){\vector(-1,0){25}}
\put(180,70){\vector(1,0){25}} \put(205,70){\line(1,0){50}}
\put(305,70){\vector(-1,0){50}} \put(330,70){\vector(-1,0){25}}
\put(180,70){\oval(200,120)[t]} \put(180,70){\circle*{5}}
\put(205,130){\vector(-1,0){25}} \put(40,55){${\bf 10}({\bf 16}_i)$}
\put(90,55){$\overline{{\bf 5}}({\bf 16}_i)$} \put(140,55){${\bf
5}({\bf 10})$} \put(190,55){$\overline{{\bf 5}}({\bf 10})$}
\put(240,55){${\bf 10}({\bf 16}_3)$} \put(290,55){$\overline{{\bf
5}}({\bf 16}_3)$} \put(160,140){${\bf 10}({\bf 45}_g)$}
\put(130,70){\line(0,-1){30}} \put(230,70){\line(0,-1){30}}
\put(175,80){$M_{10}$} \put(110,25){$\langle {\bf 1}({\bf 16}_{iH})
\rangle$} \put(210,25){$\langle \overline{{\bf 5}}({\bf 16}'_H)
\rangle$} \put(160,0){{\bf  (b)}}
\end{picture}
\end{center}

\vspace{0.2cm}

\noindent {\bf Figure 1.} The diagram in (a) shows how a tree-level
mass for ${\bf 10}_3 \overline{{\bf 5}}_i$ (shown as a blob in the
center) leads to a one-loop mass for ${\bf 10}_i \overline{{\bf
5}}_3$: i.e. the $\delta_{gi}$ elements arise
radiatively from the $C_i$ elements. The ${\bf 10}({\bf
45}_g)$ in the loop is a superheavy gauge boson. The diagram in (b)
is more detailed and shows why the loop is finite.

\vspace{1cm}

The diagram in Fig 1(a) superficially looks divergent. However, the
accidental symmetry $U(1)_F$ of the terms in Eq. (1) makes
$(M_D)_{13}$ and $(M_L)_{31}$ vanish at tree level and guarantees
that the loop is finite, as an exact calculation indeed shows. The
finiteness of this diagram is more obvious if we write it in the
form shown in Fig. 1(b). The calculation of these loops will be
discussed in section 4.

The 22 elements of $M_L$ and $M_D$ (denoted $\delta_H$) arise from
the one-Higgs-boson-loop diagram shown in Fig. 2. Whereas the
one-gauge-boson-loop shown in Fig. 1 {\it must} exist if the
tree-level masses in Eq. (3) exist, the diagram in Fig. 2 only exists
if certain couplings not needed for the tree-level masses are
present: namely, the last term in Eq. (2) ($h ({\bf 16} \; {\bf 16})
\; {\bf 10}'_H$) and a Higgs-mass term of the form ${\bf 10}_H {\bf
10}_H$. A diagram related by $SO(10)$ to the one in Fig. 2 gives a
22 element for the up-quark mass matrix $M_U$. However, if one supposes
the contributions to $(M_U)_{22}$ and $(M_D)_{22}$ from these diagrams to
be roughly comparable, then $(M_U)_{22}/(M_U)_{33}$ would be of order
$10^{-4}$ and thus at most a few percent correction to $m_c$.

\begin{center}
\begin{picture}(360,180)
\thicklines
\put(60,75){\vector(1,0){30}}
\put(90,75){\line(1,0){30}}
\put(120,75){\line(1,0){30}}
\put(180,75){\vector(-1,0){30}}
\put(180,75){\vector(1,0){30}}
\put(210,75){\line(1,0){30}}
\put(240,75){\line(1,0){30}}
\put(300,75){\vector(-1,0){30}}
\put(180,75){\oval(120,80)[t]}
\put(180,75){\vector(0,-1){15}}
\put(180,60){\line(0,-1){10}}
\put(180,115){\vector(-1,0){30}}
\put(180,115){\vector(1,0){30}}
\put(180,115){\circle*{5}}
\put(70,60){${\bf 10}({\bf 16}_2)$}
\put(130,60){${\bf 10}({\bf 16})$}
\put(195,60){${\bf 10}({\bf 16})$}
\put(250,60){$\overline{{\bf 5}}({\bf 16}_2)$}
\put(160,40){$\langle {\bf 5}({\bf 10}'_H) \rangle$}
\put(130,125){${\bf 5}({\bf 10}_H)$}
\put(190,125){$\overline{{\bf 5}}({\bf 10}_H)$}
%\put(160,0){{\bf Fig. 2}}
\end{picture}
\end{center}

\vspace{0.2cm}

\noindent
{\bf Figure 2.} A diagram showing how the 22 elements of the mass matrices can
arise
radiatively through Higgs-boson loops.

\vspace{1cm}

Before getting into a more detailed discussion of the model,
it is useful to explain how the structure of the matrices
given in Eq. (3) explains qualitatively many of the features of
the observed pattern of
masses and mixings of the quarks and leptons.

First, neglecting the $\delta$ parameters (which are of order
$10^{-2}$ because they come from one-loop diagrams) and the
parameter $\epsilon$ (which, though a tree-level effect, is
somewhat smaller than 1), one has that all the mass
matrices in Eq. (3) are of rank 1. In this approximation,
$m_b^0 \cong m^0_{\tau} \cong \sqrt{1 + |C_1|^2 + |C_2|^2}$,
where the superscript `0' denotes quantities evaluated at the GUT
scale. The relation $m^0_b \cong m^0_{\tau}$ is known to fit fairly
well in supersymmetric grand unified models with certain values of
$\tan \beta$. It works less well in non-supersymmetric grand unified
models; however, this relation will be substatially modified when
realistic symmetry breaking is taken into account in section 4.

The large (i.e. $O(1)$) off-diagonal elements $C_i$ produce large
mixing angles in the left-handed lepton sector and the right-handed
quark sector. This is because they result from mixings of
$\overline{{\bf 5}}$'s of $SU(5)$, which contain, of course,
left-handed charged leptons and right-handed down-type quarks.
Consequently, these elements produce large MNS neutrino mixing
angles, but they do not produce large CKM mixing, since CKM mixing
is of the left-handed not right-handed quarks.  This is one of the
basic ideas of so-called ``lopsided" models \cite{lopsided}.

Moreover, the present model is ``doubly lopsided" in the sense that
both $C_1$ and $C_2$ are large \cite{doublylop}. (In singly lopsided models $C_2$ is
large but not $C_1$.) This doubly lopsided structure can give a very
simple explanation of the ``bi-large" pattern of neutrino mixing
angles in the following way. Diagonalizing the charged-lepton mass matrix $M_L$ requires
$O(1)$ rotations to eliminate the elements $C_1$ and $C_2$. In
particular, it requires a rotation by $\theta_s = \tan^{-1} (C_1/C_2)$
in the 1-2 plane followed by a rotation by $\theta_a = \tan^{-1}
\sqrt{|C_1|^2 + |C_2|^2}$ in the 2-3 plane. If, as here, the neutrino mass
matrix is hierarchical, then the rotations required to diagonalize the charged lepton mass matrix $M_L$ give the dominant
contributions to the MNS matrix, which is therefore approximately of
the form

\begin{equation}
U_{MNS} = \left( \begin{array}{ccc}
c_s & s_s & 0 \\
- c_a s_s & c_a c_s & s_a \\ s_a s_s & -s_a c_s & c_a \end{array}
\right),
\end{equation}

\noindent
where $s_a \equiv \sin \theta_a$, $c_a \equiv \cos \theta_a$,
$s_s \equiv \sin \theta_s$, and $c_s \equiv \cos \theta_s$.
This is the bi-large form, with $\theta_s$ being the solar
neutrino angle and $\theta_a$ being the atmospheric angle.
When the effects of the small parameters $\epsilon$ and
the $\delta$'s are taken into account, there will be small shifts
in $U_{MNS}$, including a small non-zero $\theta_{13}$, which will be
calculated in section 5.

If one considers the effects of $\epsilon$, but still neglects the the
$\delta$'s, one sees by inspection of the mass matrices in Eq. (3)
that $m_c/m_t$ is of order $\epsilon^2$, whereas $m_s/m_b$,
$m_{\mu}/m_{\tau}$ and $V_{cb}$ are all of order $\epsilon$. This
corresponds to the experimental fact that $m_c/m_t \simeq 0.0025$,
whereas $m_s/m_b$, $m_{\mu}/m_{\tau}$, and $V_{cb}$ are given
respectively by $0.02$, $0.06$, and $0.04$. In fact, it is easy to
show from Eq. (2) that $V_{cb}^0 \simeq \frac{\epsilon}{3} \sin^2 \theta_a$ and
$(m_s/m_b)^0 \simeq \frac{\epsilon}{3} \sin \theta_a \cos \theta_a$,
so that $V_{cb} \sim m_s/m_b$, in agreement with observation. One
also easily sees that $(m_s/m_b)^0 \simeq \frac{1}{3} \epsilon \frac{C}{1 + C^2}$
and $(m_{\mu}/m_{\tau})^0 \simeq \epsilon \frac{C}{1 + C^2}$, where
$C \equiv \sqrt{|C_1|^2 + |C_2|^2}$, so that
the ratio $(m_s/m_b)^0/(m_{\mu}/m_{\tau})^0$ is
approximately $1/3$, the Georgi-Jarlskog prediction \cite{gj}.

The $u$ quark is left massless by the matrices in Eq. (3). However,
that is not a bad thing. Experimentally, $m_u/m_t$ is of order
$10^{-5}$, which is far smaller than the corresponding ratios
$m_d/m_b \sim 10^{-3}$ and $m_e/m_{\tau} \sim 0.3 \times 10^{-3}$.
Thus, if $m_d$ and $m_e$ arise at one-loop level, one would expect
$m_u$ to vanish at one-loop level.

Before analyzing the structure of the quark and lepton mass matrices
further, it is necessary to deal with the question of the pattern of
breaking of $SO(10)$ down to the Standard Model group $G_{SM} =
SU(3)_c \otimes SU(2)_L \otimes U(1)_Y$.

\section{The breaking of $SO(10)$}

If $SO(10)$ broke at a single scale all the way to the Standard Model
group $G_{SM}$, it would give the same prediction for the low energy
gauge couplings as non-supersymmetric $SU(5)$, which are known to be
unsatisfactory.  Moreover, as in non-supersymmetric $SU(5)$, the
proton lifetime would be too short.  However, it is possible to get
satisfactory gauge coupling unification and proton lifetime by
assuming a two-stage breaking with the Pati-Salam group \cite{ps}
as the intermediate symmetry:

\begin{center}
$SO(10)\stackrel{M_{G}}{\longrightarrow}SU(4)_{C}\otimes SU(2)_{L}\otimes
SU(2)_{R}\stackrel{M_{PS}}{\longrightarrow }SU(3)_{C}\otimes
SU(2)_{L}\otimes U(1)_{Y}$
\end{center}

The breaking of $SO(10)$ to the Pati-Salam group at the higher scale
$M_G$ can be done by a ${\bf 210}_H$. The breaking of the Pati-Salam
group at $M_{PS}$ is done by the VEVs of the adjoint and spinor
Higgs fields, ${\bf 45}_H$ and ${\bf 16}_{iH}$, which also
contribute to the superheavy quark and lepton masses through
couplings that appear in Eq. (2).

In running the gauge couplings between $M_G$ and $M_{PS}$, the
following matter multiplets contribute to the beta functions: (1)
The quark and lepton multiplets, ${\bf 16}_1$, ${\bf 16}_2$, ${\bf
16}_3$, ${\bf 16}$, $\overline{{\bf 16}}$, and ${\bf 10}$. Since the
masses of these multiplets are produced by coupling to the adjoint
and spinor Higgs fields and not the ${\bf 210}_H$, their splittings
are of order $M_{PS}$ and can be ignored in running between $M_G$
and $M_{PS}$. (2) The Higgs multiplets ${\bf 16}_{1H}$, ${\bf
16}_{2H}$, ${\bf 16}'_H$, ${\bf 10}_H$, ${\bf 10}'_H$, and three
adjoint Higgs multiplets. For the Higgs multiplets too, except for
the adjoints, it is assumed that the splittings are of order
$M_{PS}$ and can be neglected in running between $M_G$ and $M_{PS}$.
For the adjoints, however, we assume splittings of order $M_G$.
Under the Pati-Salam group a ${\bf 45}_H$ decomposes to $(15,1,1)_H
+ (6,2,2)_H + (1,3,1)_H + (1,1,3)_H$. We assume that the
color-singlet pieces of the adjoints get mass of order $M_G$ and the
color-non-singlet pieces get mass of order $M_{PS}$. This is not
unreasonable, since the renormalizable couplings of the adjoints to
the ${\bf 210}_H$ produce splittings of order $M_G$ between the
color-singlet and color-non-singlet pieces. (One such term is
$\langle H^{[IJKL]} \rangle H^{[IJ]} H^{[KL]}$, where the
fundamental indices $I,J,K,L$ take $SU(2)_L \otimes SU(2)_R$
values.) Of course, the whole Higgs potential must be tuned to give
the hierarchy between $M_G$ and $M_{PS}$, so different patterns of
splittings are possible. It is in order to get a value of $M_G$
large enough to be consistent with proton-decay limits, that we
assume there are three split adjoint Higgs multiplets. More such
adjoints would push $M_G$ higher. Below the scale $M_{PS}$, we
assume just the Standard Model field content with one Higgs doublet.

The results of the running are shown in Fig. 3.

\vspace{1cm}

\begin{center}
\includegraphics{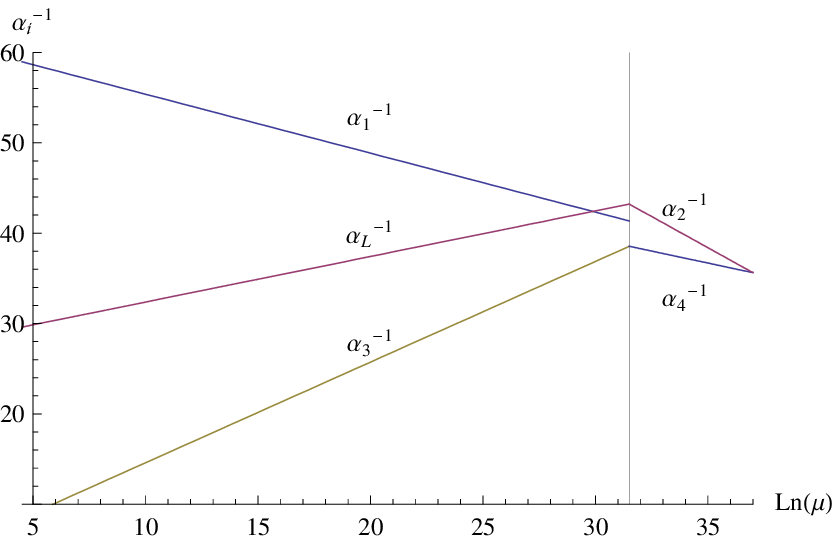}
\end{center}

\vspace{0.2cm}

\noindent {\bf Figure 3.} The running of the gauge couplings. The
Pati-Salam scale is indicated by the vertical line.

\vspace{1cm}

\noindent In the running, the input values used are $\alpha_1^{-1}
(M_Z) = 58.97$, $\alpha_2^{-1} (M_Z) = 29.61$, and $\alpha_3^{-1}
(M_Z) = 8.47$ \cite{pdg2006Raby}. The result of the running is that
$M_{PS} = 4.79 \times 10^{13}$ GeV. At $M_{PS}$ the Standard Model
couplings have the values $\alpha_1^{-1} (M_{PS}) = 41.35$,
$\alpha_2^{-1} (M_{PS}) = 43.21$, and $\alpha_3^{-1} (M_{PS}) = 38.55$.
The unification scale comes out to be $M_G = 1.17 \times 10^{16}$
GeV. and $\alpha_U^{-1} (M_G) = 35.65$. These values are consistent
with present bounds on proton decay. (In this model, the dominant
contribution to proton decay comes from dimension-6 operators
produced by the exchange of gauge bosons of mass $M_G$. The
Pati-Salam gauge bosons do not give proton decay.) The values of the
gauge couplings plotted in Fig. 3 are used in the running of the
quark and lepton masses and mixing angles in section 5.

One consequence of the fact that $SO(10)$ is broken down to the
Pati-Salam group at a high scale is that it makes more natural the
assumption being made in this model that $\langle {\bf 45}_H \rangle
\propto B-L$.  In the original supersymmetric version of the model
\cite{bb2002} this assumption was justified by the fact that a ${\bf
45}_H$ whose VEV is proportional to $B-L$ is needed to implement to
Dimopoulos-Wilczek mechanism (or ``missing VEV mechanism") \cite{dw}
for doublet-triplet splitting. However, in a non-supersymmetric
$SO(10)$ model, that mechanism does not work --- and, in fact,
doublet-triplet splitting must be achieved through fine-tuning
\cite{abds}. The justification for the assumption made in
\cite{so10rad} that $\langle {\bf 45}_H \rangle \propto B-L$ is
therefore less clear. However, if $SO(10)$ breaks to the Pati-Salam
group at the high scale $M_G$ in such a way that only the $(1,3,1)_H
+ (1,1,3)_H$ in the adjoints have the large mass $M_G$, as assumed,
then the residual $(15,1,1)_H$ naturally obtains a VEV in the $B-L$
direction (and, in fact can do no other without breaking the
Standard Model group at a superheavy scale).

\section{Modifications to the mass matrices}

In this section the implications for the mass matrices of the
breaking of $SO(10)$ down to the Pati-Salam group will be analyzed.

In Eq. (3) the term $M_{10} ({\bf 10} \; {\bf 10})$, as written,
involves an explicit mass. This was the assumption made in
\cite{so10rad}. It is also possible, however, and just as simple, to
suppose that this mass arises from the VEV of some Higgs fields that
breaks $SO(10)$. For example the term may come from $({\bf 10} \;
{\bf 10}) \langle {\bf 45}_H \rangle \langle {\bf 45}_H \rangle
/M_{PS}$ or from $({\bf 10} \; {\bf 10})  \langle {\bf 54}_H
\rangle$. If $M_{10}$ reflects the breaking of $SU(5)$ then it is a
matrix that gives different values when acting on the down quarks
and on the charged leptons in the $SO(10)$ ${\bf 10}$ of fermions.
Call its value for the leptons $M$ and for the down quarks $M'$. One
result of this splitting is that the entries $C_i$ are no longer the
same in the mass matrices $M_L$ and $M_D$.  If one assumes, as
before, that the quantities $\sqrt{1 + (\Sigma_i c_i  \langle {\bf
16}_{iH} \rangle/M_{10})^2}$ are well approximated by 1, then one
can write the $C_i$ contributions to the mass matrices as $C_i$ for
$M_L$ and $f C_i$ for $M_D$, where $f = M/M'$. This
$SO(10)$-breaking effect is, as it were, optional. However, it is
quite important for fitting $m_b^0/m_{\tau}^0$, which otherwise
would be predicted to be 1. The effects that will now be described
are unavoidable consequences of the breaking of $SO(10)$.

$SO(10)$-breaking and $SU(5)$-breaking effects come into the loop
contributions $\delta_{gi}$ and $\delta_H$ in several ways. Consider
$\delta_H$ first. If one examines the diagram in Fig. 2 closely, one
finds that its contribution to $M_D$ involves loops with scalars
that can be either color triplets or color singlets, whereas its
contributions to $M_L$ involve only color-triplet scalars.
When $SO(10)$ breaks to the Pati-Salam group, the degeneracy between
these two types of scalars is badly broken, which means that
one cannot assume that
Fig. 2 gives the same contribution to the two matrices. We therefore
introduce a factor $f_H$ into the 22 element of $M_L$ to reflect
this fact.

The case of the elements $\delta_{gi}$ requires a more involved
discussion. First, it must be noted that the vacuum expectation
values $\langle {\bf 1}({\bf 16}_{iH}) \rangle$ and $\langle {\bf
45}_H \rangle$ break the Pati-Salam group, and therefore must be no
larger than $M_{PS}$. Moreover, the masses $M_{10}$ and $M_{16}$
cannot be too much larger than these VEVs, since otherwise the
entries $C_i$ and $\epsilon$ would be too small. Consequently, one
can assume that all the superheavy fermion masses are much lighter
than the scale $M_G$. This has implications for the loop diagrams in
Fig. 1. Some of those diagrams contain gauge bosons whose mass is of
order $M_{PS}$ and others contain gauge bosons whose mass is of
order $M_G$. Because the fermions in those loops are much lighter
than $M_G$, as just argued, the loops with $O(M_G)$ gauge bosons are
suppressed relative to the loops with $O(M_{PS})$ gauge bosons by a
large factor and are therefore negligible. To see what this implies,
one must look in more detail at the diagrams in Fig. 1 to see how
they contribute to $M_L$ and $M_D$.

In Fig. 4(a), is shown the contribution to $M_L$. In this diagram
there are three possible values of the color index $a$, (or,
equivalently, of the pair of color indices $bc$ on the superheavy
gauge boson $X^{bc}$). For any of these values, the gauge boson
converts a left-handed charged lepton into a left-handed down-type
quark at one vertex, and a left-handed charged anti-lepton into a
left-handed down-type anti-quark at the other vertex. That shows
that the gauge boson is one of those of the Pati-Salam group
$SU(4)_c$, which make such transitions. (The Pati-Salam multiplet
$({\bf 4}, {\bf 2}, {\bf 1})$ unifies left-handed leptons and
quarks, while the multiplet $(\overline{{\bf 4}}, {\bf 1}, {\bf 2})$
unifies left-handed anti-leptons and anti-quarks.) For all three
values of $a$, therefore, the loop in Fig. 3(a) contains only gauge
bosons whose mass is of order $M_{PS}$.

In Figs. 4(b) and 4(c) are shown the diagrams that contribute to
$M_D$. In Fig. 4(b), the external quark has a color index of fixed
value $a$, which determines uniquely the values of the color indices
on the virtual gauge boson $X^{bc}$. This gauge boson is (as in Fig.
4(a)) one of those of the Pati-Salam group $SU(4)_c$, as can be seen
from the fact that it converts left-handed down quarks into
left-handed charged leptons. In Fig. 4(c) there are two choices for
the color index $c$ on the gauge boson $X^{1c}$, since it is only
required to be different from $a$. This gauge boson, however, is
obviously {\it not} one of those of the Pati-Salam group, since it
converts a left-handed quark into a left-handed anti-quark (and a
left-handed lepton into a left-handed anti-lepton), which are not
unified together in the multiplets of the Pati-Salam group. Thus the
gauge boson in Fig. 4(c) has mass of order $M_G \gg M_{PS}$. The
diagram in Fig. 4(c) is thus highly suppressed. From these
considerations, one sees that the contribution to $M_L$ has a factor
3 relative to the contribution to $M_D$ because of the color
degeneracy in the loop.

\begin{center}
\begin{picture}(360,140)
\thicklines \put(60,60){\vector(1,0){30}}
\put(90,60){\line(1,0){30}} \put(120,60){\vector(1,0){30}}
\put(150,60){\line(1,0){60}} \put(240,60){\vector(-1,0){30}}
\put(240,60){\line(1,0){30}} \put(300,60){\vector(-1,0){30}}
\put(180,60){\oval(120,80)[t]} \put(180,60){\circle*{20}}
\put(210,100){\vector(-1,0){30}} \put(75,45){$\psi^{12}_{(i)}$}
\put(135,45){$\psi_{(i)a}$} \put(205,45){$\psi^{2a}_{(3)}$}
\put(265,45){$\psi_{(3)1}$} \put(75,30){$=\ell^+_{(i)}$}
\put(135,30){$=\overline{d}_{(i)}$} \put(205,30){$=d_{(3)}$}
\put(265,30){$=\ell^-_{(3)}$} \put(175,110){$X^{bc}$}
\put(160,0){{\bf Fig. 4(a)}} \put(168,40){$\langle H_2 \rangle$}
\end{picture}
\end{center}

\begin{center}
\begin{picture}(360,140)
\thicklines \put(60,60){\vector(1,0){30}}
\put(90,60){\line(1,0){30}} \put(120,60){\vector(1,0){30}}
\put(150,60){\line(1,0){60}} \put(240,60){\vector(-1,0){30}}
\put(240,60){\line(1,0){30}} \put(300,60){\vector(-1,0){30}}
\put(180,60){\oval(120,80)[t]} \put(180,60){\circle*{20}}
\put(210,100){\vector(-1,0){30}} \put(75,45){$\psi^{2a}_{(i)}$}
\put(135,45){$\psi_{(i)1}$} \put(205,45){$\psi^{12}_{(3)}$}
\put(265,45){$\psi_{(3)a}$} \put(75,30){$=d_{(i)}$}
\put(135,30){$=\ell^-_{(i)}$} \put(205,30){$=\ell^+_{(3)}$}
\put(265,30){$=\overline{d}_{(3)}$} \put(175,110){$X^{bc}$}
\put(160,0){{\bf Fig. 4(b)}} \put(168,40){$\langle H_2 \rangle$}
\end{picture}
\end{center}
\begin{center}
\begin{picture}(360,140)
\thicklines \put(60,60){\vector(1,0){30}}
\put(90,60){\line(1,0){30}} \put(120,60){\vector(1,0){30}}
\put(150,60){\line(1,0){60}} \put(240,60){\vector(-1,0){30}}
\put(240,60){\line(1,0){30}} \put(300,60){\vector(-1,0){30}}
\put(180,60){\oval(120,80)[t]} \put(180,60){\circle*{20}}
\put(210,100){\vector(-1,0){30}} \put(75,45){$\psi^{2a}_{(i)}$}
\put(135,45){$\psi_{(i)b}$} \put(205,45){$\psi^{2b}_{(3)}$}
\put(265,45){$\psi_{(3)a}$} \put(75,30){$=d_{(i)}$}
\put(135,30){$=\overline{d}_{(i)}$} \put(205,30){$=d_{(3)}$}
\put(265,30){$=\overline{d}_{(3)}$} \put(175,110){$X^{1c}$}
\put(160,0){{\bf Fig. 4(c)}} \put(168,40){$\langle H_2 \rangle$}
\end{picture}
\end{center}

\vspace{0.2cm}

\noindent {\bf Fig. 4} Gauge-boson-loop contribution to mass matrices. Subscripts in
parentheses are family labels. $a,b,c$ are $SU(3)_c$ indices. $1,2$ are $SU(2)_L$
indices.

\vspace{1cm}

The gauge-boson-loop integrals can be written in a simple form if one makes the same approximation as before, namely
$\sqrt{1 + (\Sigma_i c_i  \langle {\bf 16}_{iH} \rangle/M_{10})^2}
\cong 1$.  In that case the gauge-boson-loop contributions to
$M_L$ and $M_D$ are given by

\begin{equation}
\begin{array}{cl}
(M_L)_{3i}^{gb \ell} = & 3 \left( \frac{3 \alpha_U}{16 \pi^2} \right)
C_i \frac{\ln x}{x-1},  \\
(M_D)_{i3}^{gb \ell} = & \left( \frac{3 \alpha_U}{16 \pi^2} \right) (f C_i) \frac{\ln x'}{x'-1}.
\end{array}
\end{equation}

\noindent
Here $x \equiv (M_g/M)^2$ and $x' \equiv (M_g/M')^2$, where $M_g$ is the mass of the colored
``Pati-Salam" gauge bosons in Fig. 3(a) and 3(b). Recalling that $f = M/M'$, one can rewrite these as

\begin{equation}
\begin{array}{l}
(M_L)_{3i}^{gb \ell} = 3 \left( \frac{3
\alpha_U}{16 \pi^2} \right) C_i \left(
\frac{M}{M_g} \right) F(x),  \\
(M_D)_{i3}^{gb \ell} = \left( \frac{3 \alpha_U}{16 \pi^2} \right) C_i \left(
\frac{M}{M_g} \right) F(x'),
\end{array}
\end{equation}

\noindent where $F(x) \equiv x^{1/2} \ln x/(x-1)$.  It happens that
the function $F(x)$ is very slowly varying for arguments of order 1.
For example, $F(1 + y) = 1 - \frac{1}{24} y^2 + \frac{1}{24} y^3 + ... $,
and $F(2) = F(\frac{1}{2}) = 0.98$.  Thus to a very good
approximation one may write

\begin{equation}
\begin{array}{l}
(M_L)_{3i}^{gb \ell} = 3 C_i \delta \\
(M_D)_{i3}^{gb \ell} = C_i \delta.
\end{array}
\end{equation}

\noindent The mass matrices that result are

\begin{equation}
\begin{array}{ll}
M_U = \left( \begin{array}{ccc} 0 & 0 & 0 \\
0 & 0 & \frac{\epsilon}{3} \\ 0 & - \frac{\epsilon}{3} & 1
\end{array} \right) \; m_U, \;\;\; & M_D = \left( \begin{array}{ccc}
0 & 0 & C_1 \delta \\ 0 & \delta_H & \frac{\epsilon}{3} +
C_2 \delta \\
f C_1 & f C_2 - \frac{\epsilon}{3} & 1 \end{array} \right) \; m_D, \\ & \\
M_N = \left( \begin{array}{ccc} 0 & 0 & 0 \\
0 & 0 & - \epsilon \\ 0 & \epsilon & 1 \end{array} \right) \; m_U,
\;\;\; & M_L = \left( \begin{array}{ccc}
0 & 0 & C_1 \\ 0 & f_H \delta_H & C_2 - \epsilon \\
3 C_1 \delta & \epsilon + 3 C_2 \delta & 1 \end{array} \right) \;
m_D,
\end{array}
\end{equation}

\section{Fitting the fermion masses and mixings}

The forms of the mass matrices in Eq. (8) are those given
by the model at the scale $M_{PS}$, since that is the mass of
the superheavy fields that are integrated out to give these matrices.
Below $M_{PS}$, the model reduces to the Standard Model. One can
therefore use the renormalization group equations (RGEs) of the
one-Higgs-doublet Standard Model to run the measured quark and
lepton masses and mixings from low scales up to $M_{PS}$ and then
fit the results using the forms in Eq. (8).

Running from $M_Z$ to $M_{PS}$ is done using the one-loop Standard
Model RGEs given in \cite{Barger}. The input values of the quark and
lepton masses at $M_Z$, shown in Table I, are taken from
\cite{Xing}, and were computed using updated Particle Data Group
values.
\[
\begin{tabular}{|l|l|}
\hline
$m_{u}$ & 1.27$\pm $0.50 MeV \\ \hline
$m_s/m_d$ & 19.9$\pm $0.8 \\ \hline
$m_{s}$ & 55$\pm $16 MeV \\ \hline
$m_{c}$ & 0.619$\pm $0.084 GeV \\ \hline
$m_{b}$ & 2.89$\pm $0.09 GeV \\ \hline
$m_{t}$ & 171.7$\pm $3 GeV \\ \hline
$m_{e}$ & 0.486 570 161$\pm $0.000 000 042 MeV \\ \hline
$m_{\mu }$ & 102.718 135 9$\pm $0.000 00 92 MeV \\ \hline
$m_{\tau }$ & 1746.24$\pm $0.20 MeV \\ \hline
\end{tabular}
\]
\begin{center}
{\bf Table I.}
\end{center}

\vspace{0.2cm}

 \noindent The input values of the CKM angles are
taken from \cite{pdg2004}: $s_{12}=0.2243\pm 0.0016$,
$s_{23}=0.0413\pm 0.0015$, and $s_{13}=0.0037\pm 0.0005.$
The leptonic angles are taken from \cite{pdg2006Kayser}:
$\theta_{sol} = 33.9^{\circ} \pm 2.4^{\circ}$,
$\theta_{atm} = 45^{\circ} \pm 10^{\circ}$. The quark
and lepton masses at $M_{PS}$ that result from the running are shown
in Table II.

\[
\begin{tabular}{|l|l|l|}
\hline & mass with error bar & fractional error($\%$) \\ \hline
$m_{u}$ & 0.571$\pm $0.24 MeV & $42$ \\ \hline $m_s/m_d$ & 18.9$\pm
$0.8 & 4.23 \\ \hline $m_{s}$ & 25.387$\pm $8.0 MeV & 31 \\ \hline
$m_{c}$ & 0.278$\pm $0.042 GeV & 15 \\ \hline $m_{b}$ & 1.186$\pm
$0.05 GeV & 4.2 \\ \hline $m_{t}$ & 86.926$\pm $4 GeV & 4.6 \\
\hline $m_{e}$ & 0.488848231$\pm $0.000 000 042 MeV & $10^{-5}$ \\
\hline $m_{\mu }$ & 103.199 06 11 $\pm $0.000 00 92 MeV & $10^{-5}$ \\
\hline $m_{\tau }$ & 1754.46$\pm $0.20 MeV & $10^{-4}$ \\ \hline
\end{tabular}
\]
\begin{center}
{\bf Table II.}
\end{center}

\vspace{0.2cm}

\noindent The quark mixing angles at the Pati-Salam scale are,
$s_{12}=0.2243\pm 0.0016$, $s_{23}=0.0464\pm 0.0015$ and
$s_{13}=0.0041\pm 0.0005.$

Note that we fit $m_d/m_s$, which is relatively well-known from
current algebra, rather than fitting $m_d$ itself. For several
quantities, namely the charged lepton masses, the mass of the $u$
quark, and the atmospheric neutrino angle, we will add a ``theoretical
error bar" to the experimental error bar in doing the $\chi^2$ fit.
In the case of the charged leptons, we add a $1 \%$ fractional error
to their masses, simply because we do not expect the forms in Eq.
(8) to be more accurate than that. (We have made approximations of
that order in deriving them.)

In the case of $m_u$, we add a theoretical error because the mass
matrices we are using to do the fit include only tree-level and
one-loop effects. A two-loop contribution to the 11 element of $M_U$
would be expected to be roughly of order $\left( \frac{1}{16 \pi^2}
\right)^2 m_t^0 \sim 3.5$ MeV. Thus we take the prediction of the
model to be that $m_u^0 = 0 \pm 3.5$ MeV. In other words, the theoretical
error for $m_u$ is about 600$\%$ of the actual value of $m_u$, so it
has no effect on the fitting.

In the case of the atmospheric neutrino angle, there is an inherent
uncertainty in the prediction of this model, due to the Majorana
mass matrix of the right-handed neutrinos $M_R$ being unknown and
not predicted by the model. Because the Dirac neutrino mass matrix
$M_N$ is (to one-loop order) given by the form in Eq. (8), which has
vanishing first row and column, it follows that the light neutrino
mass matrix, given by the usual type-I see-saw formula
$M_{\nu} = - M_N^T M_R^{-1} M_N$, also has vanishing
first row and column. Thus, the unitary matrix $U_{\nu}$ required to
diagonalize $M_{\nu}$ is simply a rotation by some angle
$\theta_{\nu}$ in the 23 plane. From the form of $M_N$, one expects
generically that $\theta_{\nu} = O(\epsilon)$. Thus the mixing
matrix of the neutrinos is given by

\begin{equation}
U_{MNS} = U_L U_{\nu}^{\dag} = U_L \left( \begin{array}{ccc}
1 & 0 & 0 \\ 0 & \cos \theta_{\nu} & - \sin \theta_{\nu} \\
0 & \sin \theta_{\nu} & \cos \theta_{\nu} \end{array} \right),
\end{equation}

\noindent where $U_L$ is the unitary rotation of the left-handed
charged leptons required to diagonalize $M_L$. Consequently,

\begin{equation}
\begin{array}{ccl}
\sin \theta_{atm} & = &\cos \theta_{\nu} \sin \theta_{atm}^L -
\sin \theta_{\nu} \cos \theta_{sol}^L \cos \theta_{atm}^L, \\
& = & \sin \theta_{atm}^L - O(\epsilon) \\
\sin \theta_{13} & = & \cos \theta_{\nu} \sin \theta_{13}^L - \sin \theta_{\nu}
\sin \theta_{sol}^L, \\
\sin \theta_{sol} & = & \sin \theta_{sol}^L,
\end{array}
\end{equation}

\noindent where $\sin \theta_{atm}^L$, $\sin \theta_{sol}^L$,  and
$\sin \theta_{13}^L$ are the 23, 12, and 13 elements of $U_L$
respectively. In performing the fit to the data, we take as the
neutrino mixings predicted by the model $\sin \theta_{sol}^L$, $\sin
\theta_{atm}^L$, and $\sin \theta_{13}^L$, i.e. the angles obtained
from diagonalizing the charged lepton mass matrix. These are what
are given under ``model" for these quantities in Table III. However,
we take account of the unknown $\theta_{\nu}$ by including in the
``experimental error" for $\sin \theta_{atm}$ in Table III a
``theoretical error" corresponding to an angle of $\epsilon$ = 0.19
rad.

In fitting, one has to take into account that the parameters
appearing in Eq. (8) are, in general, complex. Because of the
freedom to redefine the phases of the quark and lepton fields, most
of the phases can be ``rotated away" from the low energy theory. We
will consider two cases. If the group theoretical factors denoted
$f$ and $f_H$ are real, then there are two physical phases in the
mass matrices of Eq. (8), which one can take to be phases of the
parameters $\epsilon$ and $\delta_H$. We will denote these by
$\theta_{\epsilon}$ and $\theta_H$. If $f$ and $f_H$ are complex,
then there are two additional phases, which we will denote $\theta'$
and $\theta_{fH}$. The phase $\theta'$ comes into the subleading
terms of the 23 and 32 elements of $M_D$ and $M_L$. The phase
$\theta_{fH}$ comes only into the 22 element of $M_L$. The phase
$\theta'$ has only a small effect on the fit, and $\theta_{fH}$ has
almost none.

Altogether, then, we fit using 9 real parameters ($M_U$, $M_D$,
$C_1$, $C_2$, $\epsilon$, $\delta$, $\delta_H$, $f$, $f_H$) and two
or four phases ($\theta_{\epsilon}$, $\theta_H$, and if $f$ and
$f_H$ are complex then also $\theta'$ and $\theta_{fH}$). With these
we fit sixteen quantities: the 9 masses of the quarks and leptons
(excluding the neutrino masses, which depend on the unknown $M_R$),
the 3 CKM angles, the 1 CKM phase, and the 3 neutrino mixing angles.
\[
\begin{tabular}{|l|l|l|l|l|}
\hline & model (at $M_{PS}$) & expt. (at $M_{PS}) $ &  off ($\%$) & expt. error$^*$ ($\%$) \\
\hline $m_e$ & 0.0004900 & 0.0004888 & 0.027 & 1.0$^*$ \\
\hline $m_{\mu}$ & 0.1031 & 0.1032 & -0.13 & 1.0$^*$ \\
\hline $m_{\tau}$ & 1.756 & 1.754 & 0.07 & 1.0$^*$ \\
\hline $m_u$ & 0 & 0.000571 & 100 & 600$^*$ \\
\hline $m_c$ & 0.342 & 0.278 & 23.0 & 15.1 \\
\hline $m_t$ & 87.24 & 86.93 & 0.36 & 4.6 \\
\hline $m_s/m_d$ & 18.68 & 18.90 & -1.14 & 4.23 \\
\hline $m_s$ & 0.0358 & 0.0254 & 40.8 & 31.5 \\
\hline $m_b$ & 1.17 & 1.186 & -1.29 & 4.22 \\
\hline $\frac{m_e}{m_{\mu}}/\frac{m_d}{m_s}$ & 0.0886 & 0.0895 &
-0.99 & \\
\hline $m_b/m_{\tau}$ & 0.667 & 0.676 & -1.35 & \\
\hline $V_{us}$ & 0.2243 & 0.2243 & 0.002 & 0.71 \\
\hline $V_{cb}$ & 0.0456 & 0.0463 & -1.51 & 3.24 \\
\hline $|V_{ub}|$ & 0.00368 & 0.00432 & -14.8 & 11.6 \\
\hline $\delta_{13}$ & 0.887 & 0.995 & -10.8 & 24.12 \\
\hline $\sin \theta_{sol}$ & 0.518 & 0.559 & -7.33 & 7.51 \\
\hline $\sin \theta_{atm}$ & 0.891 & 0.707 & 26.1 & 28$^*$ \\
\hline $\sin \theta_{13}$ & 0.014 & $<0.178$ & & \\
\hline $\chi^2$ & 7.2 & & & \\ \hline
\end{tabular}
\]
\begin{center}
{\bf Table III.}
\end{center}

The results of the fit assuming $f$ and $f_H$ are real are given in
Table III. The asterisks in the ``experimental error" column are
reminders that for certain entries a ``theoretical error" is
included, as explained above. The masses are all in GeV. The CKM
phase $\delta_{13}$ is in radians. One notes that most quantities
are fit excellently. The least good fits are to $m_c$, $m_s$, and
$|V_{ub}|$. Considering that 11 parameters are fitting 16
quantities, the $\chi^2$ of 7.2 is quite reasonable.

The parameter values for this fit are $\epsilon = 0.189$, $C_1 =
1.03$, $C_2 = -1.51$, $f = 0.566$, $f_H = 0.208$, $16 \pi^2 \delta =
2.22$, $16 \pi^2 \delta_H = 2.66$, $\theta_{\epsilon} = 1.52$ rad,
$\theta_H = 0.514$ rad. Note that all these quantities are of order
one. In other words, no small dimensionless parameters are needed to
fit the quark and lepton mass hierarchies in this model. The scales
called $m_U$ and $m_D$ in Eq. (8) are given respectively by 86.9 GeV
and 0.79 GeV. The large ratio of these scales is not explained by
the structure of the model or by symmetry, and presumably comes from
the details of the sector that breaks the weak interactions.

\[
\begin{tabular}{|l|l|l|l|l|}
\hline & model (at $M_{PS}$) & expt. (at $M_{PS}) $ &  off ($\%$) & expt. error$^*$ ($\%$) \\
\hline $m_e$ & 0.0004888 & 0.0004888 & -0.012 & 1.0$^*$ \\
\hline $m_{\mu}$ & 0.1032 & 0.1032 & -0.01 & 1.0$^*$ \\
\hline $m_{\tau}$ & 1.755 & 1.754 & 0.02 & 1.0$^*$ \\
\hline $m_u$ & 0 & 0.000571 & 100 & 600$^*$ \\
\hline $m_c$ & 0.317 & 0.278 & 14.14 & 15.1 \\
\hline $m_t$ & 87.22 & 86.93 & 0.33 & 4.6 \\
\hline $m_s/m_d$ & 18.44 & 18.90 & -2.39 & 4.23 \\
\hline $m_s$ & 0.0346 & 0.0254 & 36.17 & 31.5 \\
\hline $m_b$ & 1.17 & 1.186 & -1.67 & 4.22 \\
\hline $\frac{m_e}{m_{\mu}}/\frac{m_d}{m_s}$ & 0.0874 & 0.0895 &
-2.39 & \\
\hline $m_b/m_{\tau}$ & 0.665 & 0.676 & -1.7 & \\
\hline $V_{us}$ & 0.2243 & 0.2243 & -0.018 & 0.71 \\
\hline $V_{cb}$ & 0.0458 & 0.0463 & -1.13 & 3.24 \\
\hline $|V_{ub}|$ & 0.00382 & 0.00432 & -11.5 & 11.6 \\
\hline $\delta_{13}$ & 0.889 & 0.995 & -10.6 & 24.12 \\
\hline $\sin \theta_{sol}$ & 0.499 & 0.559 & -10.7 & 7.51 \\
\hline $\sin \theta_{atm}$ & 0.895 & 0.707 & 26.7 & 28$^*$ \\
\hline $\sin \theta_{13}$ & 0.015 & $<0.178$ & & \\
\hline $\chi^2$ & 6.0 & & & \\ \hline
\end{tabular}
\]
\begin{center}
{\bf Table IV.}
\end{center}

The results of the fit assuming $f$ and $f_H$ are complex are given
in Table IV. The parameter values for the fit in Table IV are
$\epsilon = 0.182$, $C_1 = 0.997$, $C_2 = -1.60$, $f = 0.573$, $f_H
= 0.224$, $16 \pi^2 \delta = 2.16$, $16 \pi^2 \delta_H = 3.22$,
$\theta_{\epsilon} = -0.554$ rad, $\theta_H = -1.56$ rad.

Comparison of Tables III and IV shows that the inclusion of the
phases $\theta'$ and $\theta_{fH}$ makes very little difference to
the fits. This is not surprising, since $\theta'$ appears only on
subleading terms in the mass matrices, and $\theta_{fH}$ appears on
the very small entry $f_H$. For the two fits, the values of the real
parameters hardly changes. The phase angles $\theta_{\epsilon}$ and
$\theta_H$ both change by -2.07 rad, but that is essentially due to
a rephasing: a shift in these two phases by a certain amount can be
compensated by a shift in $\theta'$ together with a change in the
phase of two small subleading terms. In other words, what is really
changing a lot between the fits in Tables III and IV is the phase
$\theta'$ (which is, of course, zero for the fit in Table III). What
this shows is that the fit is hardly affected by large changes in
$\theta'$.

In this model there is a relation between the atmospheric
angle and $\theta_{13}$, which is given in Eq. (10). Using the best-fit
values given in Table III, Eq. (10) yields

\begin{equation}
\begin{array}{ccl}
|\sin \theta_{atm}| & = & |\cos \theta_{\nu} (0.891) -
\sin \theta_{\nu} (0.396)|, \\
|\sin \theta_{13}| & = & |\cos \theta_{\nu} (-0.014) + \sin
\theta_{\nu} (0.518)|.
\end{array}
\end{equation}

\noindent If the parameter $\sin \theta_{\nu}$ were a real number, these
equations would give a prediction for $\theta_{13}$ in terms of $\theta_{atm}$.
For values of the atmospheric angle near maximal
mixing, i.e. $\theta_{atm} \cong \pi/4$, the prediction for the 13
angle would be approximately given by

\begin{equation}
|\sin \theta_{13}| \cong 0.160 - 0.72(\sin \theta_{atm} - 1/\sqrt{2}).
\end{equation}

\noindent However, in fact, the parameter $\sin \theta_{\nu}$ can be
complex. Therefore, the smaller of the two values that is obtained
for $|\sin \theta_{13}|$ by solving Eq. (11) with real $\sin
\theta_{\nu}$ is a {\it lower bound}. So, if the atmospheric mixing
angle is near maximum, there is a lower bound on $\sin \theta_{13}$
given by Eq. (12).

\section{Conclusions}

The model that we have studied here is the first predictive grand
unified model with a radiative fermion mass hierarchy. In a number
of ways, it is as economical as a model of quark and lepton masses
can be. The masses and mixings of the second and third families come
from only three effective Yukawa operators, shown in Eq. (1). These
operators account for many features of the light fermion spectrum:
(1) the fact that $V_{cb}$ is of the same order as $m_s/m_b$ and
$m_{\mu}/ m_{\tau}$; (2) the fact that $m_c/m_t$ is much smaller
than those ratios; (3) the largeness of the atmospheric and solar
neutrino angles; (4) the smallness of the 13 angle; (5) the rough
equality of $m_b^0$ and $m_{\tau}^0$; and (6) the Georgi-Jarlskog
factor of about 1/3 between $m_s/m_b$ and $m_{\mu}/m_{\tau}$. The
masses and mixings of the first family (except for the solar
neutrino angle) come from loop diagrams. It is remarkable that one
of these loop diagrams (Fig. 1) is present automatically, whereas
the other (Fig. 2) requires only a single additional Yukawa term to
be postulated.

It is striking that no small parameters are needed in this model to
account for the dramatic hierarchies in the quark and lepton masses.
The model yields a definite relation between the atmospheric angle
and the angle $\theta_{13}$.

\end{document}